\documentclass[apj,twocolappendix, numberedappendixn]{emulateapj}
\usepackage[colorlinks,citecolor=blue]{hyperref}
\usepackage[colorlinks,citecolor=blue]{hyperref}
\usepackage{textcomp}
\usepackage{subfigure}
\usepackage{verbatim}
\usepackage{bm}% bold math
\usepackage{hyperref}
\usepackage{amsmath}
\usepackage{graphicx}
\usepackage{epstopdf}
\usepackage{amssymb}
\usepackage{extarrows}
\usepackage{color}
\usepackage{CJK}
\usepackage{cancel}
\usepackage{ulem}
\usepackage[utf8]{inputenc}
\usepackage{url}
\usepackage{float}
\usepackage{lineno}

\newcommand{\ba}{\begin{eqnarray}}
\newcommand{\ea}{\end{eqnarray}}
\newcommand{\be}{\begin{equation}}
\newcommand{\ee}{\end{equation}}
\newcommand{\gr}{\mathrm{GR}}

\newcommand{\au}{\mathrm{AU}}

\newcommand{\tb}{\mathrm{TB}}
\newcommand{\td}{\mathrm{TD}}

\begin{document}
\title{Planetary Desert around Compact Binaries: Dynamical Instability Triggered by Resonance-Induced Eccentricity Excitation}
\author{Bin Liu$^{1,2}$, Dong Lai$^{3,4}$}
\affil{$^{1}$ Institute for Astronomy, School of Physics, Zhejiang University, Hangzhou 310027 , China\\
$^{2}$ Center for Cosmology and Computational Astrophysics, Institute for Advanced Study in Physics, Zhejiang University, Hangzhou 310027, China\\
$^{3}$ Tsung-Dao Lee Institute, Shanghai Jiao Tong University, Shanghai 200240, China\\
$^{4}$ Department of Astronomy, Center for Astrophysics and Planetary Science, Cornell University, Ithaca, NY 14853, USA
}

\begin{abstract}
Compact binaries with orbital periods shorter than about 7 days show an absence of transiting planets,
a feature known as the ``circumbinary planet desert". The physical mechanism behind this desert remains unclear.
We investigate its origin by simulating the long-term dynamics of multi-planet circumbinary systems with evolving inner binaries.
Our simulations are based on the single-averaged secular equations that average only over the binary orbital period
and fully incorporate planet-planet interactions.
When an eccentric binary decays via tides, an outer planet can be captured into resonance advection in eccentricity,
a state in which its apsidal precession locks with that of the binary, driving extreme eccentricity growth.
While such growth can occur in a binary-single planet system, the parameter space is limited and may not necessarily induce instability.
In a multi-planet system, however, the excited orbit inevitably crosses those of its neighbors,
which triggers violent planet-planet scatterings and produces collisions or ejections.
Crucially, these mutual gravitational interactions amplify the ``localized" instability of a single planet into a system-wide chain reaction,
drastically reshaping the orbital architecture and potentially clearing out the inner regions of planetary systems.
Our results suggest that the resonance-induced instability provides a natural explanation for the observed circumbinary planet desert.
\end{abstract}
%\keywords{ Binary stars; Exoplanet dynamics; Star-planet interactions}

%  \rightharpoondown
%\maketitle

\section{Introduction}

Recent observations have identified over 30 planets (or candidates) orbiting stellar or brown-dwarf binaries through various detection methods,
including planet transits \citep[e.g.,][]{Doyle2011},
timing variations of eclipsing binaries \citep[e.g.,][]{Borkovits2016,Getley2017},
radial velocity variations \citep[e.g.,][]{Correia2005,Standing2023,Goldberg2023,Baycroft 2025},
microlensing \citep[e.g.,][]{Bennett2016,Kuang2022,Han2024},
and direct imaging \citep[e.g.,][]{Burgasser2010,Kuzuhara2011,Delorme2013,Bryan2016,Bonavita2017,Dupuy2018,Dupuy2023,Janson2019,Janson2021}.
Among these, 14 transiting circumbinary planets (CBPs) have been discovered by the \textit{Kepler} and \textit{TESS} missions
\citep[e.g.,][]{Orosz2012a,Orosz2012b,Orosz2019,Schwamb2013,Leung 2013,Welsh2012,Welsh2015,Martin 2015b, Kostov2020, Kostov2021,Socia2020}.
These CBPs are in nearly coplanar orbits
around their host binaries with periods $T_{\rm b} \gtrsim 7$ days, and many reside near the dynamical stability limit
\citep[e.g.,][]{Armstrong2014,Windemuth2019}.
Such a configuration is consistent with planet's formation history,
in which the circumbinary protoplanetary disk is truncated by the binary's torque,
and the planets migrate inward until stalling near the disk inner edge
\citep[e.g.,][]{Nelson2003,Kley2012,Kley2015,Rafikov2014,Thun2018,Penzlin2021}.

However, a striking discrepancy emerges for compact binaries.
Although eclipsing binaries with periods shorter than a few days are abundant in the \textit{Kepler} target list,
none have been found to host transiting planets\citep[e.g.,][]{Orosz2012a}.
The absence of planets around compact binaries with $T_{\rm b} \lesssim 7$ days is statistically significant
and cannot be attributed to observational biases, as geometric transit probability favors detection around tighter systems.
This deficit therefore points to a physical mechanism that either suppresses planet formation or destabilizes planetary orbits
around the most compact stellar binaries.

Several mechanisms have been proposed. The tight binaries themselves may have formed through high-eccentricity migration driven
by an inclined tertiary companion via the von Zeipel–Lidov–Kozai (ZLK) effect
\citep[e.g.,][]{vonZeipel1910,Mazeh1979,Lidov1962,Kozai1962,Fabrycky2007,Naoz2014,Smadar Review,Moe2018},
during which pre-existing CBPs are likely engulfed, ejected, or placed on distant, misaligned orbits
\citep[e.g.,][]{Martin2015a,Munoz2015,Hamers2016}.
Other processes include evection resonances with external perturbers \citep[e.g.,][]{Touma1998,Xu2016},
or complex orbital evolution due to tidal synchronization and magnetic braking \citep[e.g.,][]{Verbunt1981,Ivanova2003,Fleming2018}.
Alternatively, the observed ``planetary desert" could partly arise from observational concealment,
such as a reduced transit probability after binary orbital shrinkage \citep[e.g.,][]{Mogan2025},
or from the intrinsic misalignment of circumbinary planets forming in tilted disks
\citep[e.g.,][]{Doolin 2011,Martin 2017,Zanazzi 2018,Cuello 2019,Martin 2022,Martin 2025}.

In this work, we propose a dynamical mechanism that operates during the tidal decay of a stellar binary.
Recent studies of coplanar triple systems show that when the inner binary undergoes orbital decay,
a tertiary body can be captured into a state of resonance advection,
where its apsidal precession locks with that of the binary, driving extreme eccentricity growth
\citep[e.g.,][]{Liu2015a,Liu2020,Liu2022,Liu2024,Farhat2025}.
Based on this scenario, we study a system consisting of an eccentric stellar binary and multiple CBPs.
For simplicity, we adopt coplanar configurations, as expected for circumbinary systems originating from aligned disks.
As the binary experiences orbital shrinkage due to tidal dissipation, some of the planets are driven into apsidal precession resonances,
exciting their eccentricities and triggering orbital crossings.
The mutual gravitational interactions among the planets therefore become significant,
driving widespread instabilities.
We examine how this process reshapes the orbital architecture of the tidally evolved binary system.
Note that for highly inclined CBP orbits,
the inverse ZLK effect can induce long-term eccentricity and inclination oscillations \citep[e.g.,][]{Naoz 2017, Vinson 2018,de Elia 2019}.
However, it does not apply to our coplanar configurations.

This \textit{Letter} is organized as follows.
Section \ref{sec 2} reviews the physics of apsidal precession resonance capture and advection,
and demonstrates how extreme eccentricity excitation can occur during the
orbital shrinkage of inner binary.
In Section \ref{sec 3}, we show that the planet-planet scattering,
triggered by the resonant eccentricity excitation, amplifies the ``local" instability of a single planet
into global system clearing through collisions and ejections.
We summarize our conclusions in Section \ref{sec 4}.

\section{Eccentricity excitation due to apsidal precession resonance}
\label{sec 2}

We consider a hierarchical triple system in a coplanar configuration.
The inner stellar binary consists of stars with masses $m_1$ and $m_2$.
The CBP with mass $m_p=0.1M_\mathrm{J}$ and radius $R_p=0.2R_\mathrm{J}$
\footnote{
Although CBPs span a range of masses (from a few Earth masses to several Jupiter masses)
and radii (typically between 0.3 and 0.8 $R_{\rm J}$), their precise distributions remain uncertain \citep[e.g.,][]{Li2016,Welsh2018,Muller 2024}.
We adopt a fiducial planetary mass of 0.1 $M_{\rm J}$ and radius of 0.2 $R_{\rm J}$ for our calculations.
Note that larger masses or radii would enhance planet–planet scattering and collision probabilities,
further promoting dynamical clearing (see Section \ref{sec 3}).}.

orbits the center of mass of the inner binary.
The semi-major axis and eccentricity of the stellar binary are denoted by $a_b$ and $e_b$,
and those of the planetary orbit by $a_p$ and $e_p$, respectively.
The orbital angular momenta are given by
$\textbf{L}_b\equiv\mathrm{L}_b\hat{\textbf{L}}_b=\mu_b\sqrt{G m_{12}a_b(1-e_b^2)}\,\hat{\textbf{L}}_b$
and $\textbf{L}_p\equiv\mathrm{L}_p\hat{\textbf{L}}_p\simeq\mu_p\sqrt{G m_{12}a_p(1-e_p^2)}\,\hat{\textbf{L}}_p$.
where $m_{12}\equiv m_1+m_2$, $\mu_b\equiv m_1m_2/m_{12}$ and $\mu_p\simeq m_p$.

To study the long-term evolution of planetary orbits during the gradual tidal decay of the stellar binary,
we adopt the single-averaged (SA; only averaging over the inner orbital period) secular equations of motion.
In contrast to the widely used double-averaging method (averaging over both the inner and outer orbital periods)
\citep[e.g.,][]{Naoz 2013,Petrovich 2015,Liu2015b}
and Gauss's method \citep[e.g.,][]{Touma 2009}, both of which allow only angular momentum exchange between the orbits,
the SA equations of motion enable us to study dynamical planet–planet interactions in multi-planet circumbinary systems.
Specifically, the SA formalism captures both angular momentum and energy exchange among planet orbits,
thereby allowing the subsequent processes of collisions, scattering, and ejections to be modeled (see Section \ref{sec 3}).

The evolution of the inner binary itself is described by the following vector equations:
%%%%%%%%%%%%%%%%%%%%%%%%%%%%%%%%%%%%%%%%%%%%%%%%%%%%%%%%%%%%%%%%%%%%%%
\begin{eqnarray}
&&\frac{d \textbf{L}_b}{dt}=\frac{d \textbf{L}_b}{dt}\bigg|_\td~,\label{eq:Full Kozai 1}\\
&&\frac{d \mathbf{e}_b}{dt}=\frac{d \mathbf{e}_b}{dt}\bigg|_\gr
+\frac{d \mathbf{e}_b}{dt}\bigg|_\tb+\frac{d \mathbf{e}_b}{dt}\bigg|_\td~,\label{eq:Full Kozai 2}
\end{eqnarray}
%%%%%%%%%%%%%%%%%%%%%%%%%%%%%%%%%%%%%%%%%%%%%%%%%%%%%%%%%%%%%%%%%%%%%%
where the subscripts denote contributions from general relativistic (GR) precession,
precession due to the tidal bulge (TB), and tidal dissipation (TD), respectively.
We adopt the formalism of \citet{Hut1981,Leconte2010}
and parameterize the rate of stellar tidal dissipation using a weak-friction model with a constant tidal lag time
\footnote{In this framework, the viscous dissipation timescale $t_V$ is the primary source of uncertainty.
Although $t_V$ varies by $1–2$ orders of magnitude across stellar types, we adopt $t_V=1$yr as our fiducial value.
Tidal dissipation is included for only $m_1$ and assume the rotation rate is pseudo-synchronous.
The effective tidal quality factor is $Q=4kGm_1 t_V/[3(1+2k)^2 R_1^3 n_b]$
with $k=0.014$, where $n_b=(Gm_{12}/a_b^3)^{1/2}$ is the mean motion of the inner binary.
To test the sensitivity of our results to the assumed decay rate, we have performed additional simulations with $t_V=50$yr;
a detailed comparison is presented in the Appendix \ref{Appendix A}.}.
The explicit forms of these terms can be found in \citet{Eggleton2001,Fabrycky2007}.
Note that the evolution of the inner binary can also be affected by the tertiary companion
through mechanisms such as the ZLK effect.
However, because the planetary mass is much smaller than the binary mass ($m_p=0.1M_\mathrm{J}\ll m_{12}$),
the gravitational feedback of the planet on the binary evolution is negligible and can be safely ignored
\footnote{
We do not consider the tidal evolution of the planetary orbit in the present study.
Even when the eccentricity of the CBP's orbit becomes large and the pericenter distance approaches within a few times the binary semi-major axis,
the separation between the planet and either star remains sufficiently large that tidal distortion induced on the planet is negligible
(see Figure \ref{fig:parameter space}).
}.

For the CBP's orbit, the long-term evolution is governed by the perturbing potential induced by the inner stellar binary $\langle\Phi\rangle$:
%%%%%%%%%%%%%%%%%%%%%%%%%%%%%%%%%%%%%%%%%%%%%%%%%%%%%%%%%%%%%%%%%%%%%%
\be\label{eq:third body}
\mu_p \frac{d^2\mathbf{r}_p}{dt^2}=\nabla_{\mathbf{r}_p}\bigg(\frac{Gm_{12}m_p}{r_p}\bigg)-
m_p\nabla_{\mathbf{r}_p}\langle\Phi\rangle,
\ee
%%%%%%%%%%%%%%%%%%%%%%%%%%%%%%%%%%%%%%%%%%%%%%%%%%%%%%%%%%%%%%%%%%%%%%
where $\mathbf{r}_p\equiv\mathrm{r}_p\hat{\mathbf{r}}_p$ denotes the position vector of the planet
relative to the center of mass of the inner bodies.
The averaged potential $\langle\Phi\rangle$ includes the quadrupole and octupole Newtonian gravitational potentials of the inner binary
\citep[e.g.,][]{Liu2015b,Liu2018}.
Equations (\ref{eq:Full Kozai 1})-(\ref{eq:third body}) fully determine the dynamics of the planetary orbit under the SA approximation.

We further consider systems with multiple planets.
We develop a ``Ring$+$Nbody" code and incorporate the mutual Newtonian gravitational potential
between the planets into $\langle\Phi\rangle$ in Equation (\ref{eq:third body}).
For the long evolution timescales considered here ($\gtrsim10^7$ years),
this method offers a substantial speedup over a full N-body integration
while preserving the energy exchange needed to model planet–planet scattering, collisions, and ejections.

For a coplanar hierarchical triple, a secular eccentricity excitation can occur through apsidal precession resonance.
This resonance operates when the precession rate of the inner binary periastron, $\dot\varpi_b$, matches that of the outer (planetary) orbit, $\dot\varpi_p$
\citep[e.g.,][]{Liu2015a,Liu2020,Liu2022}:
%%%%%%%%%%%%%%%%%%%%%%%%%%%%%%%%%%%%%%%%%%%%%%%%%%%%%%%%%%%%%%%%%%%%%%%%%%%%%%%%%%
\be\label{eq:APR}
\dot\varpi_b\simeq\dot\varpi_p.
\ee
%%%%%%%%%%%%%%%%%%%%%%%%%%%%%%%%%%%%%%%%%%%%%%%%%%%%%%%%%%%%%%%%%%%%%%%%%%%%%%%%%%
When this condition is satisfied, efficient eccentricity exchange between the two orbits becomes possible
\footnote{
Note that significant inclination excitation can occur for initially near-coplanar triples \citep[e.g.,][]{Li2014},
but the required conditions ($m_1>m_p\gg m_2$ and large eccentricities) are not satisfied for the systems studied here.
So, we do not consider this effect in our study.
}.

For our system,
the inner binary's precession rate is dominated by general relativistic (1PN) pericenter advance and the non-dissipative tidal bulge contribution
%%%%%%%%%%%%%%%%%%%%%%%%%%%%%%%%%%%%%%%%%%%%%%%%%%%%%%%%%%%%%%%%%%%%%%
\be\label{eq: precession rate inner}
\dot\varpi_b=\dot\varpi_\gr+\dot\varpi_\tb,
\ee
%%%%%%%%%%%%%%%%%%%%%%%%%%%%%%%%%%%%%%%%%%%%%%%%%%%%%%%%%%%%%%%%%%%%%%
where
%%%%%%%%%%%%%%%%%%%%%%%%%%%%%%%%%%%%%%%%%%%%%%%%%%%%%%%%%%%%%%%%%%%%%%
\be
\dot\varpi_\gr=\frac{3G^{3/2} m_{12}^{3/2}}{c^2a_b^{5/2} (1-e_b^2)},
\ee
%%%%%%%%%%%%%%%%%%%%%%%%%%%%%%%%%%%%%%%%%%%%%%%%%%%%%%%%%%%%%%%%%%%%%%
and the tidal bulge contribution (for $m_1$) takes the form \citep[e.g.,][]{Liu2015b}
%%%%%%%%%%%%%%%%%%%%%%%%%%%%%%%%%%%%%%%%%%%%%%%%%%%%%%%%%%%%%%%%%%%%%%
\be
\dot\varpi_\tb=\frac{15}{8}\sqrt{\frac{Gm_{12}}{a_b^3}}\bigg(\frac{m_2}{m_1}\bigg)
\bigg(\frac{R_1}{a_b}\bigg)^5k\frac{8+12e_b^2+e_b^4}{(1-e_b^2)^5},
\ee
%%%%%%%%%%%%%%%%%%%%%%%%%%%%%%%%%%%%%%%%%%%%%%%%%%%%%%%%%%%%%%%%%%%%%%
where $k=0.014$ and $R_1\simeq0.85m_1^{0.67}R_\odot$ are the classical apsidal motion constant and the radius of $m_1$, respectively
\citep[e.g.,][]{Demircan1991}.
For simplicity, we neglect the precession induced by the rotational bulge - this is of similar order as
$\dot\varpi_\tb$ for pseudo-synchronous rotation.

The planetary orbit precession rate arises from the secular gravitational perturbation of the binary in quadrupole order \citep[e.g.,][]{Liu2015b}
%%%%%%%%%%%%%%%%%%%%%%%%%%%%%%%%%%%%%%%%%%%%%%%%%%%%%%%%%%%%%%%%%%%%%%
\be\label{eq: precession rate outer}
\dot\varpi_p=\frac{3}{4}\bigg(\frac{G\mu_b m_p a_b^2}{a_p^3}\bigg)\frac{1}{L_p(1-e_p^2)^{3/2}}.
\ee
%%%%%%%%%%%%%%%%%%%%%%%%%%%%%%%%%%%%%%%%%%%%%%%%%%%%%%%%%%%%%%%%%%%%%%

\begin{figure}
\centering
\begin{tabular}{cccc}
\includegraphics[width=8.5cm]{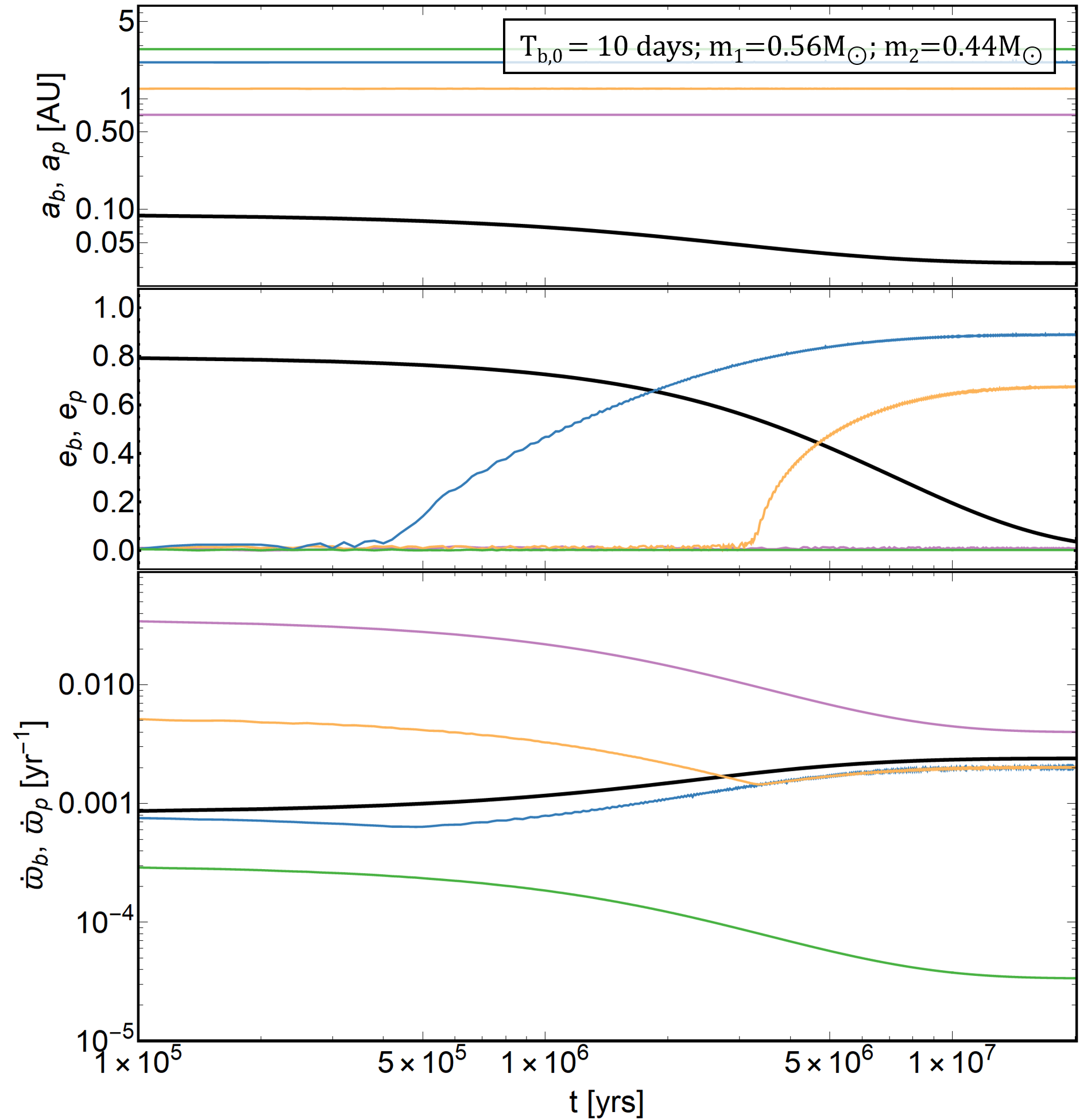}
\end{tabular}
\caption{Evolution of a stellar binary (black) with four (non-interacting) circumbinary planets in a coplanar configuration.
The masses of the inner stellar binary are $m_1=0.56M_\odot$ and $m_2=0.46M_\odot$.
The initial semimajor axis and eccentricity of the inner binary are $a_{b,0}=0.09\au$ and $e_{b,0}=0.8$.
The outer orbits, ordered by increasing distance, have initial semimajor axes $a_p=0.72\au$, $1.24\au$, $2.13\au$ and $2.81\au$.
All four begin with eccentricity $e_{p,0}=0.001$.
The longitudes of pericenter for both the binary and the planetary orbits are initialized to $\varpi_{b,0}=\varpi_{p,0}=0$.
Each binary–planet system is evolved independently (i.e., without planet–planet interactions)
by integrating the single-averaged (SA) secular equations (Equations \ref{eq:Full Kozai 1}-\ref{eq:third body}),
where the tidal dissipation is included in the binary evolution (top and middle panels).
The bottom panel shows the corresponding apsidal precession rates computed from
Equations (\ref{eq: precession rate inner}) and (\ref{eq: precession rate outer}).
}
\label{fig:SA evolution}
\end{figure}

Recent studies have shown that if the inner binary undergoes adiabatic orbital decay
(e.g., via GW emission or tidal dissipation), the tertiary can experience extreme eccentricity excitation
\citep[e.g.,][]{Liu2024,Farhat2025}.
In particular, it can be captured into a state of resonance advection,
during which the two precession rates are in a $1:1$ commensurability
until the binary circularization is reached.
This resonance-locking phenomenon leads to a continuous growth of the outer orbit eccentricity $e_p$.

Three principal criteria must be satisfied for successful resonance capture and advection \citep[][]{Liu2024}:
\textit{i)} Adiabatic evolution: The timescale of binary orbital decay must be much longer than the characteristic secular timescale
$|\dot a_b/a_b|^{-1}\gg|\dot\varpi_p|^{-1}$.
\textit{ii)} Direction of resonance crossing: The system must encounter the resonance from the regime
$\dot\varpi_p>\dot\varpi_b$ towards $\dot\varpi_p<\dot\varpi_b$.
\textit{iii)} Small initial outer binary eccentricity:
The initial $e_{p,0}$ must be less than about $15(m_1-m_2)a_b e_b(4+3e_b^2)/(8m_{12}a_p)$
at the moment of $\dot\varpi_b\simeq\dot\varpi_p$.

To illustrate the resonance advection, Figure \ref{fig:SA evolution} depicts the long-term evolution of
the CBP during the orbital decay of the stellar binary due to tidal dissipation.
We consider four planets with different initial separations, and
each is initially placed outside the critical stable semi-major axis $a_{p,\mathrm{c}}$
given by the empirical formula of \citet{Holman1999}:
%%%%%%%%%%%%%%%%%%%%%%%%%%%%%%%%%%%%%%%%%%%%%%%%%%%%%%%%%%%%%%%%%%%%%%
\ba\label{eq: aoutc}
&&a_{p,\mathrm{c}}=a_b(1.6+5.1e_b-2.22e_b^2+4.12\mu_\mathrm{c}-4.27e_b\mu_\mathrm{c}\nonumber\\
&&~~~~~~~~-5.09\mu_\mathrm{c}^2+4.61e_b^2\mu_\mathrm{c}^2),
\ea
%%%%%%%%%%%%%%%%%%%%%%%%%%%%%%%%%%%%%%%%%%%%%%%%%%%%%%%%%%%%%%%%%%%%%%
where $\mu_\mathrm{c}=m_2/m_{12}$.
The results are obtained by integrating the SA secular equations for individual triples, i.e.,
the planet-planet interactions are not included.

Driven by tidal dissipation, we see that the binary semi-major axis $a_b$ shrinks while its eccentricity $e_b$ decreases,
reaching a final separation of $a_{b,\mathrm{f}}\simeq0.03\au$ (corresponding to an orbital period of 2.2 days).
The inner binary apsidal precession rate $\dot\varpi_b$ (Equation~\ref{eq: precession rate inner}) increases monotonically,
whereas the planetary orbital precession rate $\dot\varpi_p$ (Equation~\ref{eq: precession rate outer})
decreases as the binary's gravitational potential weakens. The evolution leads to three distinct outcomes.
For initially distant planets ($a_p=2.81$ AU),
the initial precession rate $\dot\varpi_{p,0}$ is smaller than $\dot\varpi_{b,0}$;
hence the precession matching condition is never satisfied and no resonance excitation occurs.
When the planet is closer ($a_p=2.13$ AU and $a_p=1.24$ AU),
the initial configuration satisfies $\dot\varpi_{b,0} \lesssim \dot\varpi_{p,0}$.
As $\dot\varpi_b$ grows and overtakes $\dot\varpi_p$, the system is captured into resonance advection.
The two precession rates then evolve in lockstep until the inner binary is circularized,
driving a continuous rise in the planetary eccentricity $e_p$.
A comparison between the planetary pericenter distance and the critical semi-major axis $a_{p,\mathrm{c}}$ (Equation~\ref{eq: aoutc})
shows that no CPB becomes unstable during this phase.
For the innermost planet ($a_p=0.72$ AU),
$\dot\varpi_p$ remains larger than $\dot\varpi_b$ throughout the evolution, and no resonance passage occurs.
We therefore find that resonant eccentricity excitation requires that the binary semi-major axis at resonance,
$a_{b,\mathrm{res}}$ (obtained by solving Equation~\ref{eq:APR} for $a_b$)
exceed its final value $a_{b,\mathrm{f}}$ when tidal evolution ceases.
Note that this is different from the system that undergoes orbital decay due to gravitational wave emission or magnetic braking \citep[see][]{Liu2024}.
In those cases, $\dot\varpi_b$ continues to increase, allowing the tertiary to eventually satisfy the resonance condition.

\begin{figure}
\centering
\begin{tabular}{cccc}
\includegraphics[width=8.5cm]{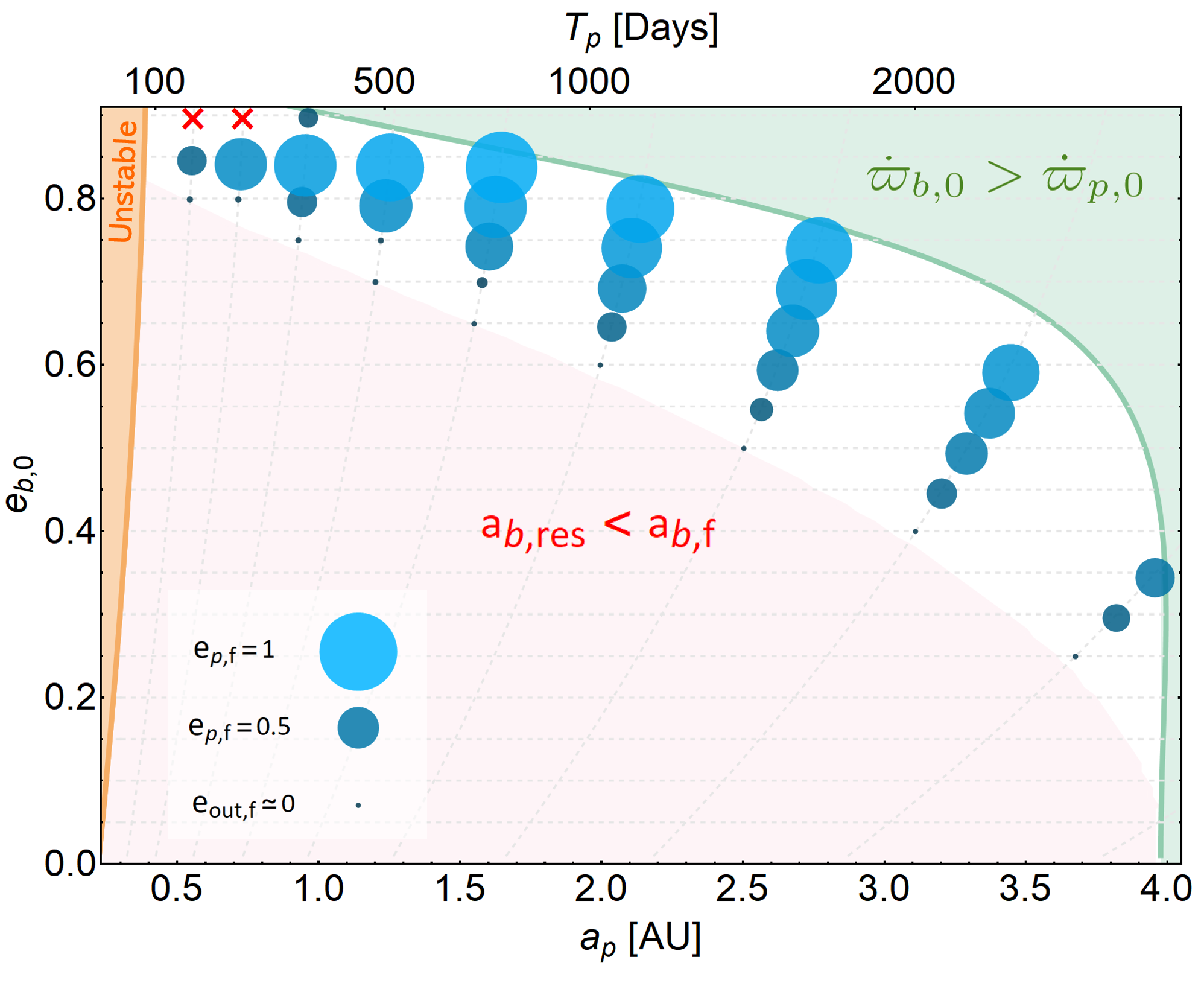}
\end{tabular}
\caption{Parameter space in the $a_p-e_{b,0}$ plane showing the regime
for resonance capture and advection during the orbital decay of the stellar binary.
We consider coplanar configurations for all simulations presented in this figure.
The orbital period $T_p$ for planets at different distances is labeled at the top.
We choose the same stellar binary as in Figure \ref{fig:SA evolution}, varying only the initial binary eccentricity $e_{b,0}$.
The orange region marks systems that are dynamically unstable according to Equation (\ref{eq: aoutc})
by considering initially circular planetary orbits.
The initial orbits of the planets are spaced according to the mutual Hill instability criterion with $K_{\mathrm{c}} = 10$
in Equation (\ref{eq: Hill radius}).
The green region corresponds to systems in which the initial apsidal precession rate of the inner binary
exceeds the precession rate of the planetary orbit.
The pink region represents systems for which the binary semimajor axis never shrinks to the value
required to trigger apsidal precession resonance before tidal decay ceases.
Systems in the green and pink regions would not experience resonance during the binary orbital decay.
For each planetary orbit, we integrate the single-averaged (SA) secular equations until
tidal evolution ends and record the final eccentricity $e_{p,\mathrm{f}}$,
whose value is represented by the size of the circles.
The red crosses indicate orbits that become unbound in the numerical integration—either
the semimajor axis turns negative or the eccentricity exceeds unity.
}
\label{fig:parameter space}
\end{figure}

To characterize the parameter dependence of the resonance excitation,
Figure \ref{fig:parameter space} maps the parameter space in the $a_p-e_{b,0}$ plane where the eccentricity
of a planetary orbit $e_p$ can be excited.
For a given binary, the initial orbital spacing of the planets follows the Hill instability criterion
%%%%%%%%%%%%%%%%%%%%%%%%%%%%%%%%%%%%%%%%%%%%%%%%%%%%%%%%%%%%%%%%%%%%%%
\be\label{eq: Hill radius}
a_{p,j}-a_{p,i}=K_{\mathrm{c}}\bigg(\frac{a_{p,i}+a_{p,j}}{2}\bigg)\bigg(\frac{2m_p}{m_{12}}\bigg)^{1/3},
\ee
%%%%%%%%%%%%%%%%%%%%%%%%%%%%%%%%%%%%%%%%%%%%%%%%%%%%%%%%%%%%%%%%%%%%%%
where $i, j$ denote adjacent planets, and $K_{\mathrm{c}}$ is a constant.
We adopt $K_{\mathrm{c}}=10$ in this study \citep[e.g.,][]{Pu2021} and set the innermost planet's semi-major axis to
1.5 $a_{p,\mathrm{c}}$ (see Equation \ref{eq: aoutc}).

\begin{figure*}
\centering
\begin{tabular}{cccc}
\includegraphics[width=8.5cm]{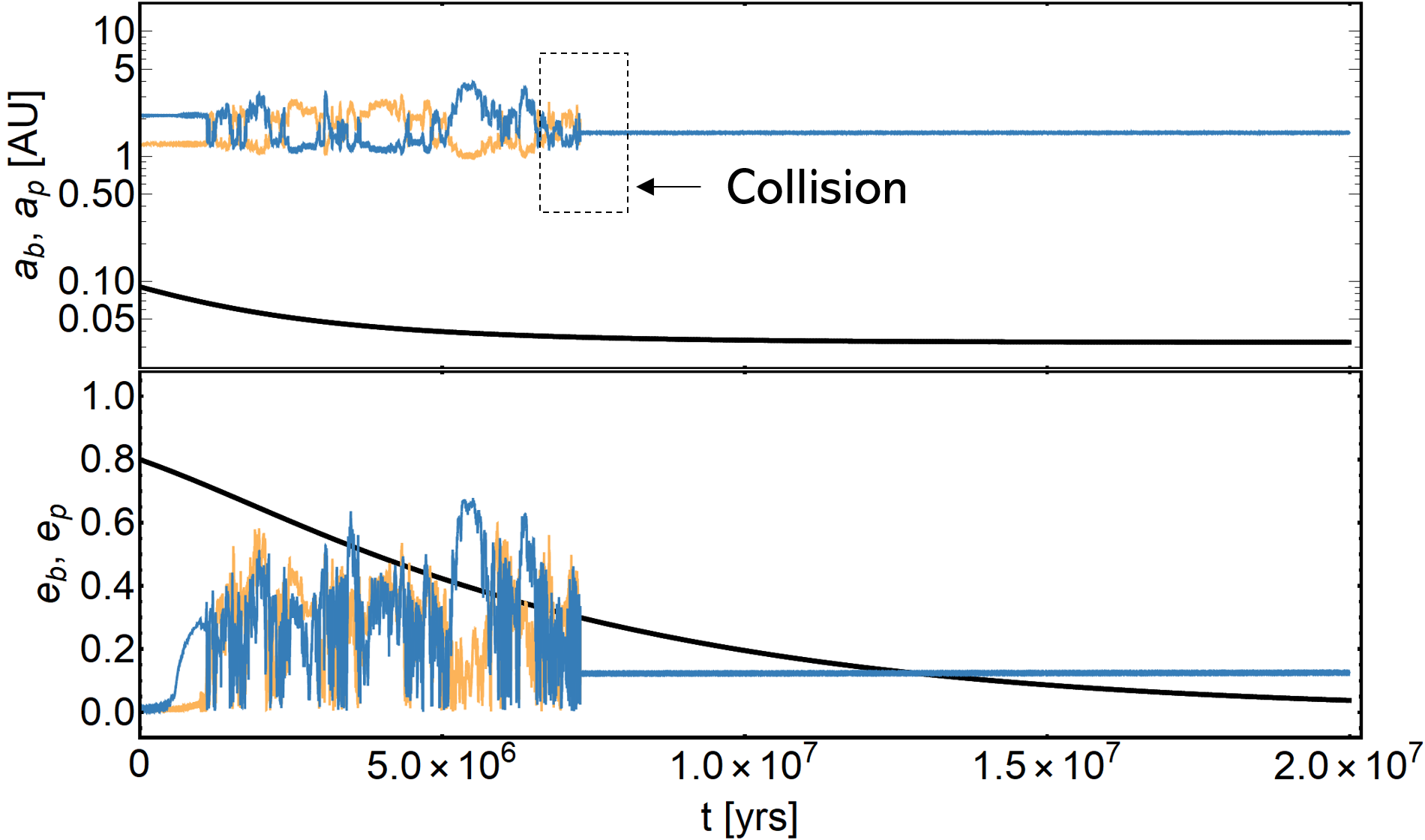}
\includegraphics[width=8.5cm]{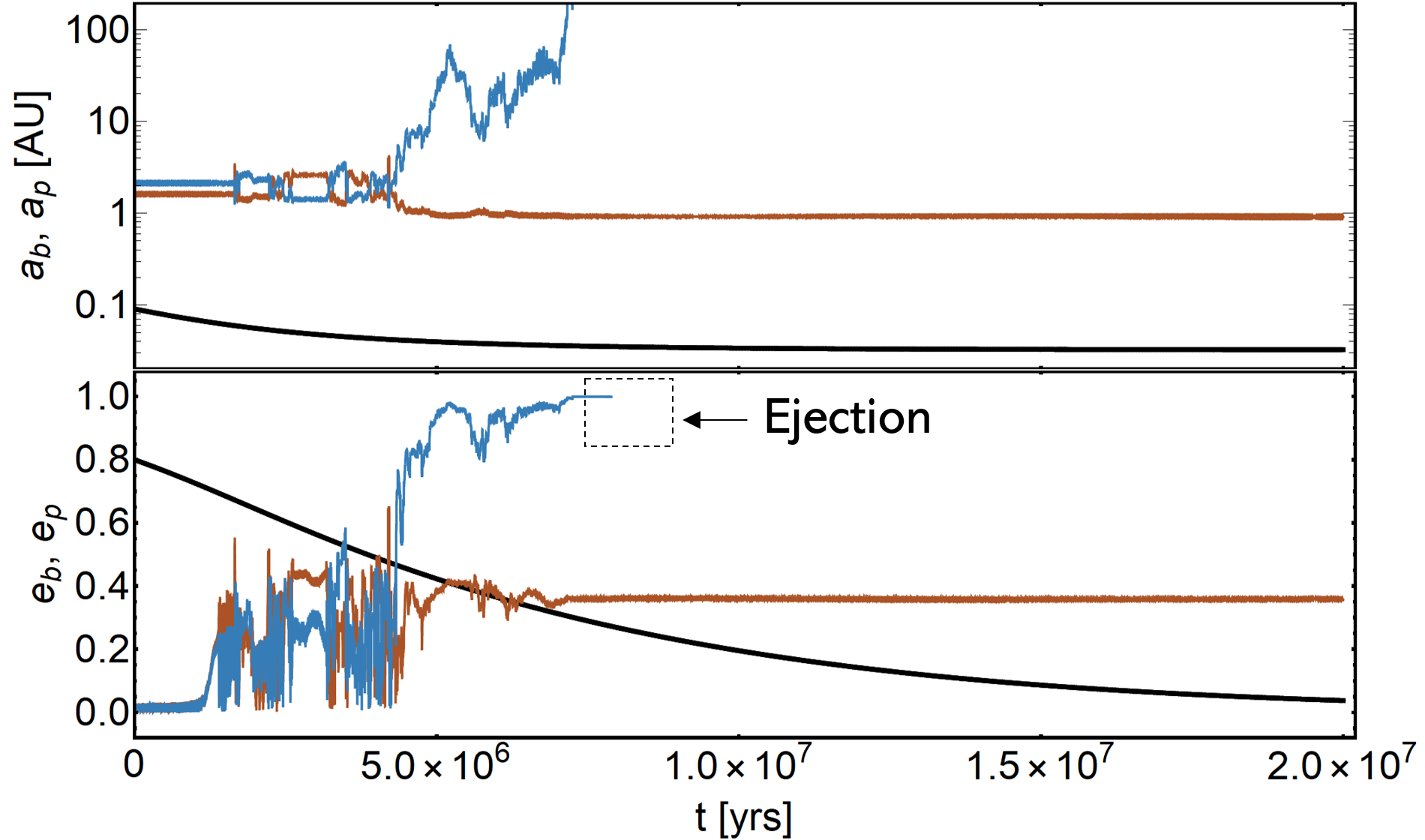}
\end{tabular}
\caption{Similar to Figure \ref{fig:SA evolution}, but including mutual gravitational interaction between two CBPs.
The initial longitude of pericenter of each planet is randomly chosen between $0$ and $2\pi$.
The evolution of the system can lead to either collision or ejection.
Left panel: Two planets with initial semimajor axes $a_p=2.13\au$ and $1.24\au$ collide into a single body.
Right panel: One planet initially located at $a_p=2.13\au$ is ejected from the system.
}
\label{fig:SA two rings evolution}
\end{figure*}

Figure \ref{fig:parameter space} reveals a finite range of planetary semi-major axes that permits $e_p-$excitation.
A planet cannot be captured into apsidal precession resonance during binary orbital decay in the green and pink regions.
First, if the planet is too close to the binary,
the resonance location $a_{b,\mathrm{res}}$ falls inside the binary’s final orbital radius $a_{b,\mathrm{f}}$
(see the evolution of the $a_p=0.72\au$ orbit in Figure \ref{fig:SA evolution}).
Second, if the planet is too distant, the initial precession rates satisfy $\dot\varpi_{b,0}>\dot\varpi_{p,0}$
(see the evolution of the $a_p=2.81\au$ orbit in Figure \ref{fig:SA evolution}).
The parameter space for systems capable of undergoing resonance advection is limited,
and becomes narrow as the initial binary eccentricity $e_{b,0}$ decreases.
This results from the competition between the tidal decay timescale and the eccentricity growth timescale.
Moreover, for a given $e_{b,0}$, planets located at larger $a_p$ can achieve higher final eccentricities.
This results from their prolonged resonance locking phase, which persists over most of the binary tidal decay,
as illustrated in Figure \ref{fig:SA evolution}.

\section{Secular Evolution and Gravitational Scattering in Multi-Planet Systems}
\label{sec 3}

\begin{figure*}
\centering
\begin{tabular}{cccc}
\includegraphics[width=8.5cm]{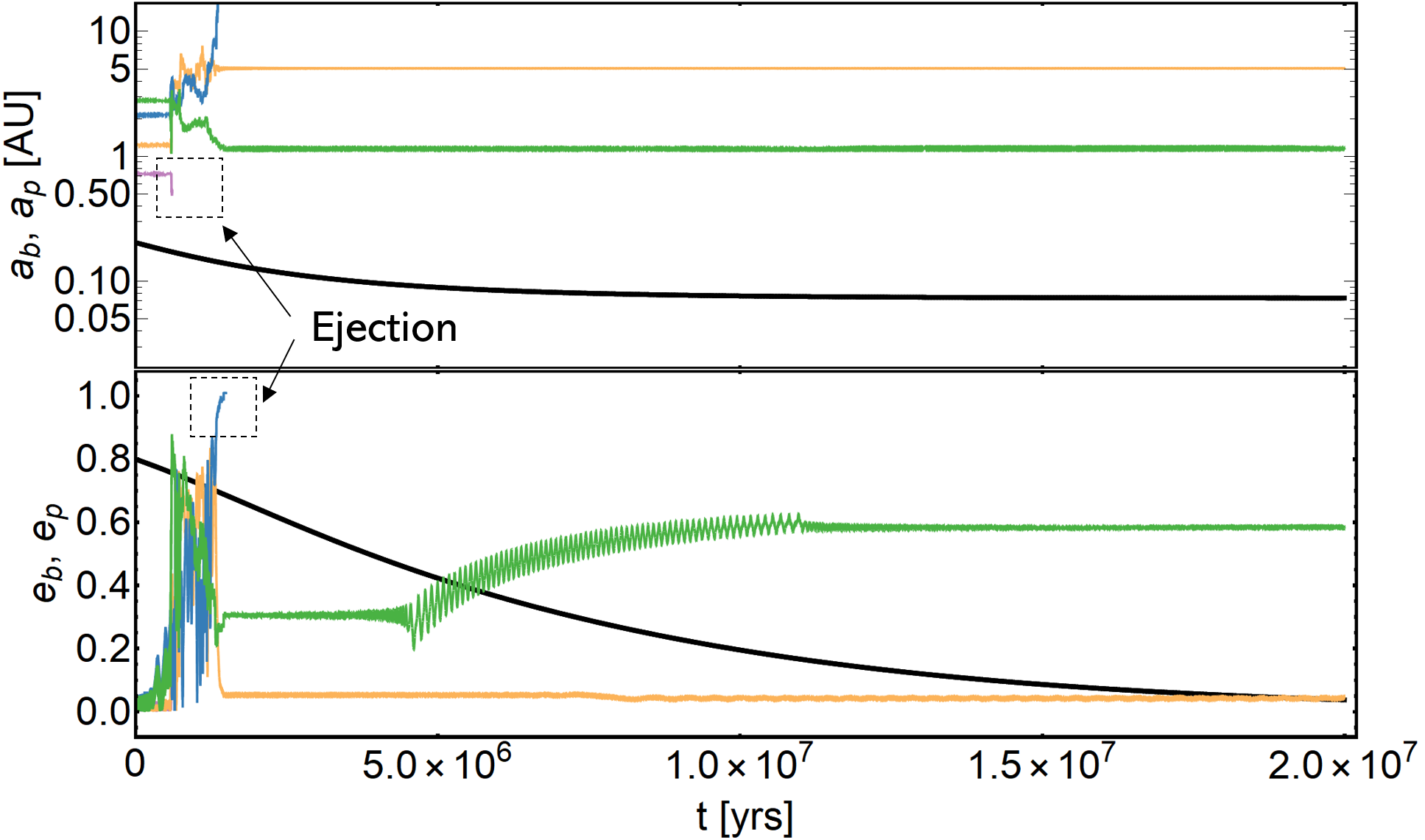}
\includegraphics[width=8.5cm]{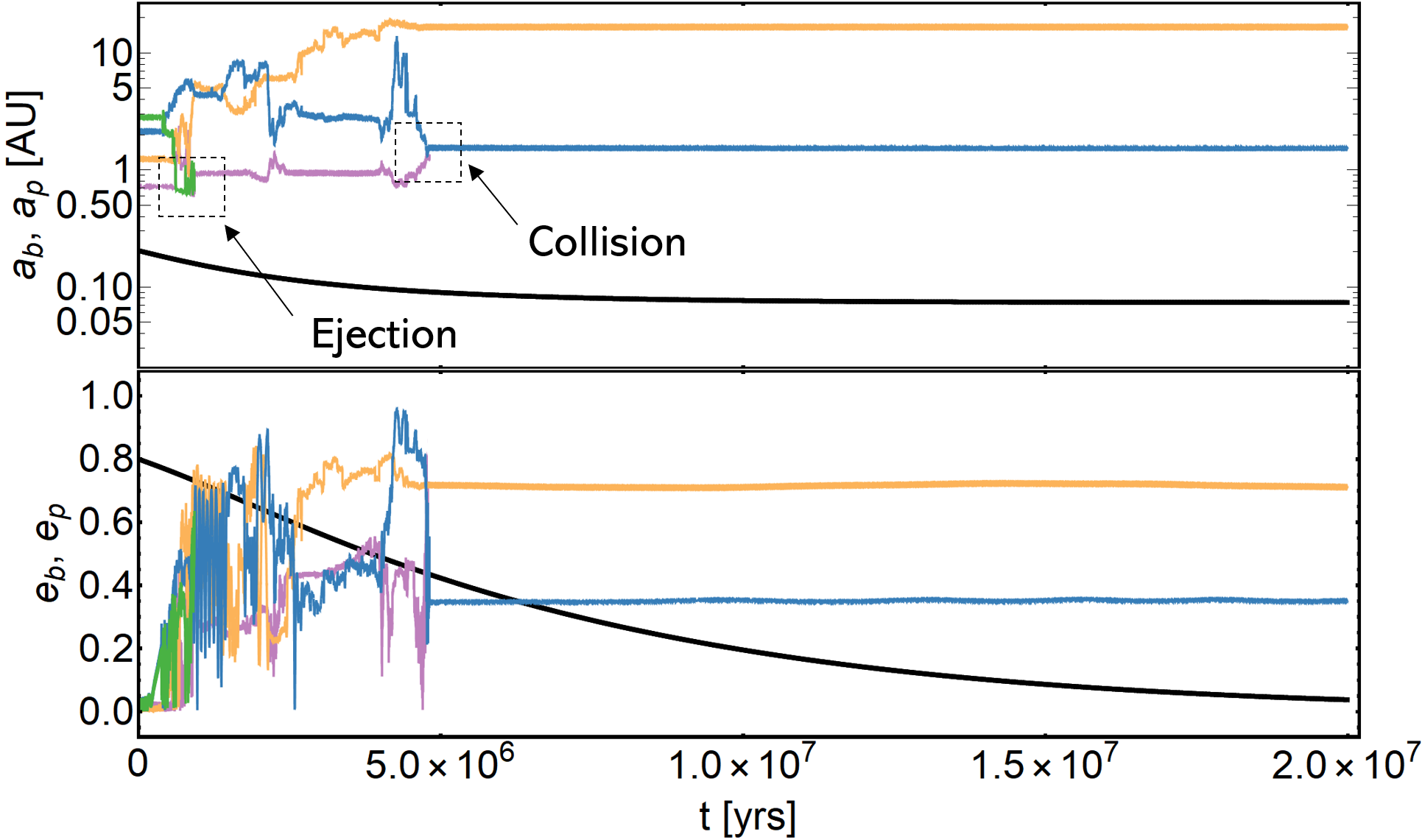}
\end{tabular}
\caption{Same system as in Figure \ref{fig:SA evolution}, but with mutual gravitational interactions among all four planets.
The initial longitude of pericenter of each CBP's orbit is randomly chosen in the range [$0-2\pi$].
The eccentricity of the planetary orbit grows significantly, even for the planet
that does not undergo resonant eccentricity-excitation in isolation.
The resulting orbital crossings trigger strong interactions between planets,
which reshape the orbital architecture and lead to either an ejection event (left panel) or a collision event (right panel).}
\label{fig:SA four rings evolution}
\end{figure*}

We now consider systems with multiple planets that are interacting with each other.
Once a planet's eccentricity $e_p$ undergoes resonant excitation,
the periastron distance of the orbit decreases sharply while the apastron distance expands.
This causes the planetary orbit to cross those of its neighbors,
triggering close encounters and gravitational scattering.
At this stage, mutual gravitational interactions among planets replace the long-term resonant perturbations from the host binary
as the dominant driver of the system dynamics.
This scattering process redistributes both angular momentum and orbital energy,
resulting in drastic changes to the planetary semi-major axes, and potentially
leading to either ejection of the planets or collisions with neighboring bodies.

\subsection{Case study: two circumbinary planets}

When two planets undergo orbital crossing and enter the regime of strong scattering,
their evolutionary outcomes include three types:

{\parindent0pt\textit{i) Ejection}}.
Previous N-body numerical experiments indicate that in systems comprising two planets on initially unstable orbits,
the averaged timescale for an ejection can be approximated as
%%%%%%%%%%%%%%%%%%%%%%%%%%%%%%%%%%%%%%%%%%%%%%%%%%%%%%%%%%%%%%%%%%%%%%
\be\label{eq: ejection time}
T_\mathrm{ej}\simeq0.06^2\bigg(\frac{m_{12}}{m_{p,i}}\bigg)^2\bigg(1+\frac{m_{p,j}}{m_{p,i}}\bigg)^4T_{p,j},
\ee
%%%%%%%%%%%%%%%%%%%%%%%%%%%%%%%%%%%%%%%%%%%%%%%%%%%%%%%%%%%%%%%%%%%%%%
where $T_{p,j}$ is the initial orbital period of $m_{p,j}$ \citep[e.g.,][]{Pu2021,Li2022}.
Equation (\ref{eq: ejection time}) implies that ejection typically requires many orbital periods.
In our simulations, an ejection event is recorded if the semi-major axis of a planet becomes negative ($a_p<0$)
or if its eccentricity exceeds unity ($e_p>1$).

{\parindent0pt\textit{ii) Collision}}.
Whether two planets collide or scatter apart depends primarily on the ``Safronov number":
%%%%%%%%%%%%%%%%%%%%%%%%%%%%%%%%%%%%%%%%%%%%%%%%%%%%%%%%%%%%%%%%%%%%%%
\be\label{eq: Safronov number}
\Theta\equiv \bigg(\frac{v_\mathrm{esc}}{v_\mathrm{orb}}\bigg)^2,
\ee
%%%%%%%%%%%%%%%%%%%%%%%%%%%%%%%%%%%%%%%%%%%%%%%%%%%%%%%%%%%%%%%%%%%%%%
where $v_\mathrm{esc}$ is the escape velocity of the planet
and $v_\mathrm{orb}$ is its orbital velocity around the star.
If the Safronov number is less than unity, a significant fraction of planetary collisions are expected.
In our simulations, a collision is identified when the separation between two planets becomes smaller than the sum of their radii, 2$R_p$.
Such an event is denoted as, for example, Orbit $i-$Orbit $j$ ($Oi-Oj$ hereafter).
After a collision, we continue to track the evolution of the collision remnant,
which has a mass equal to the sum of the colliding bodies but an unchanged radius,
and is labeled as $Oj'$.
The position and velocity of this remnant are set by momentum conservation at the instant of collision.

{\parindent0pt\textit{iii) Tidal disruption}}.
The third possible outcome involves interactions between a planet and an individual star of the host binary, such as tidal disruption.
However, because the SA equations can not resolve the precise positions of the stars along their orbits,
such a process is not considered in this study.

We explore the evolution of a two-planet system, as shown in Figure \ref{fig:SA two rings evolution}.
The initial semi-major axes of both planets lie within the parameter space for resonant excitation
identified in Figure \ref{fig:parameter space}.
As the host binary shrinks, the eccentricities of the two planets are sequentially excited.
Orbital crossing between their elongated orbits occur, triggering strong scattering.
The left panel illustrates a case where the two planets collide and merge into one body.
The right panel illustrates an outcome in which one planet is ejected from the system.

\subsection{Case study: four circumbinary planets}

\begin{figure}
\centering
\begin{tabular}{cccc}
\includegraphics[width=8.5cm]{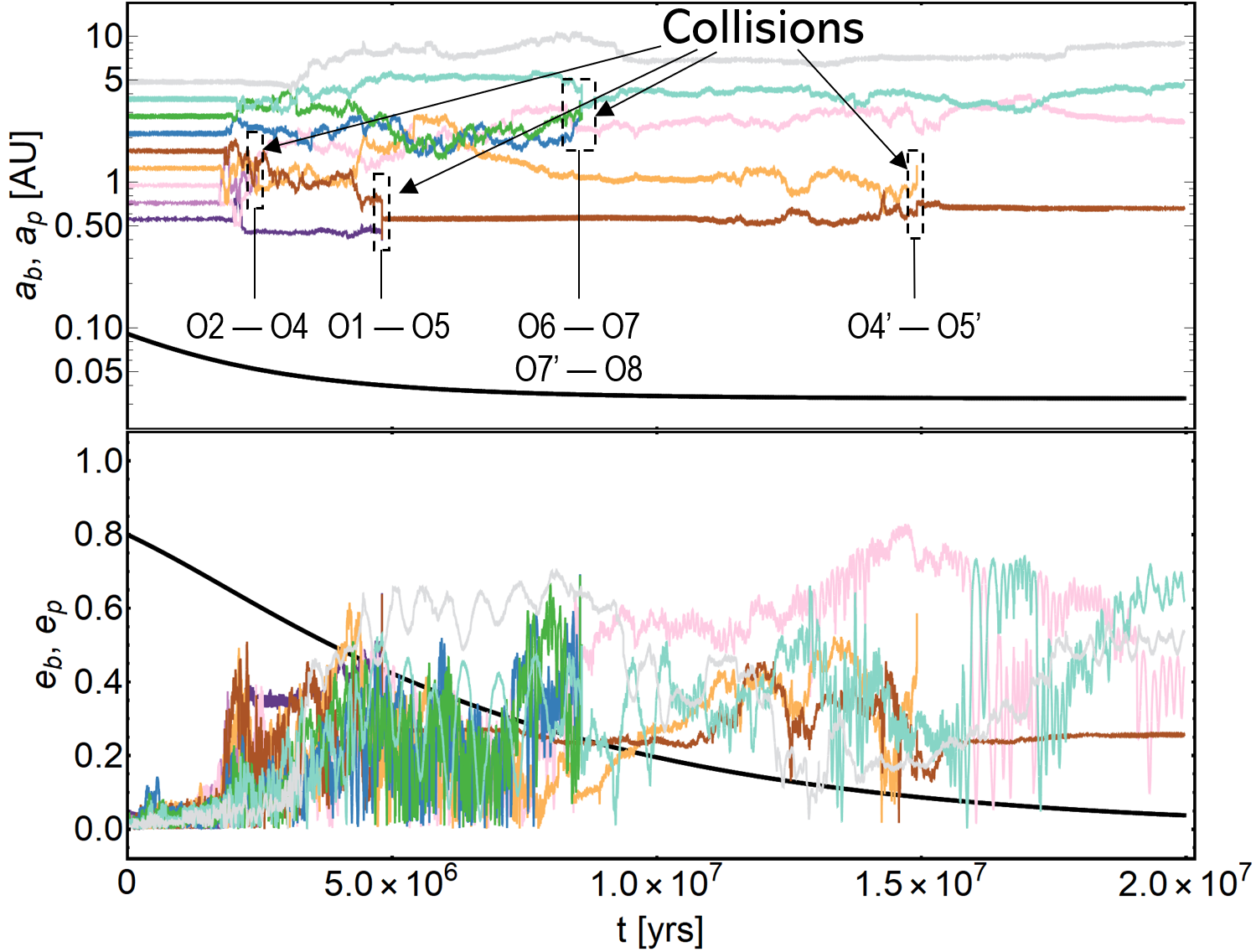}
\end{tabular}
\caption{The evolution of $9-$planet system during tidal decay of the host binary,
with mutual planet-planet gravitational interactions included. The system parameters are taken from Figure \ref{fig:parameter space},
with $e_{b,0}=0.8$ as a fiducial value:
The planets are ordered by increasing distance with semimajor axes
$a_p=0.55\au$, $0.72\au$, $0.94\au$, $1.24\au$, $1.63\au$, $2.13\au$, $2.81\au$, $3.69\au$, $4.85\au$.
Mutual interactions lead to frequent planetary collisions,
with some planets experiencing repeated collisions,
e.g., O2 and O4 merge into O4', O1 and O5 merge into O5', and then O4' and O5' collide to form O5''.
}
\label{fig:SA nine rings evolution}
\end{figure}

Figure \ref{fig:SA four rings evolution} shows the evolution of a system containing four interacting planets
\citep[e.g.,][]{Chen 2023,Chen 2024,Childs 2023,Coleman 2024},
with the initial parameters of the system identical to those in Figure \ref{fig:SA evolution}.

A key finding is that planet-planet interactions can dramatically extend the parameter space for eccentricity excitation.
Even planets that would remain stable in isolation
(e.g., those located at $a_p=0.72$AU or $2.81$AU, shown in Figures \ref{fig:SA evolution}-\ref{fig:parameter space})
are perturbed by neighboring planets whose eccentricities are resonantly excited.
These perturbations induce substantial indirect eccentricity growth,
leading to orbital crossings that allow the planets to approach closely to each other and produce mutual collisions or ejections.
We see that the evolution of the planets is strongly chaotic, evidenced by the sensitive
dependence of the outcomes (left vs. right panels) on the initial conditions.

\subsection{Case study: nine circumbinary planets}

To evaluate how a ``localized" instability triggered by resonance propagates throughout the planetary system
via gravitational coupling among the planets,
here we consider an extreme example of nine planets,
and carry out Monte Carlo simulations with random initial logitude of pericenter to obtain statistical results.

In Figure \ref{fig:SA nine rings evolution}, the nine planets are initially spaced according to
Equation (\ref{eq: Hill radius}) with $K_{\mathrm{c}}=10$.
During the tidal decay of the binary, only a subset of planets, concentrated in the region $0.9\au\lesssim a_p\lesssim2.2\au$,
are directly excited by the apsidal precession resonance.
We see that the orbital evolution becomes increasingly complex as the number of planets increases.
Nevertheless, the scattering events induced by mutual gravitational interactions quickly spread throughout the inner planetary system.

\begin{figure}
\centering
\begin{tabular}{cccc}
\includegraphics[width=8.5cm]{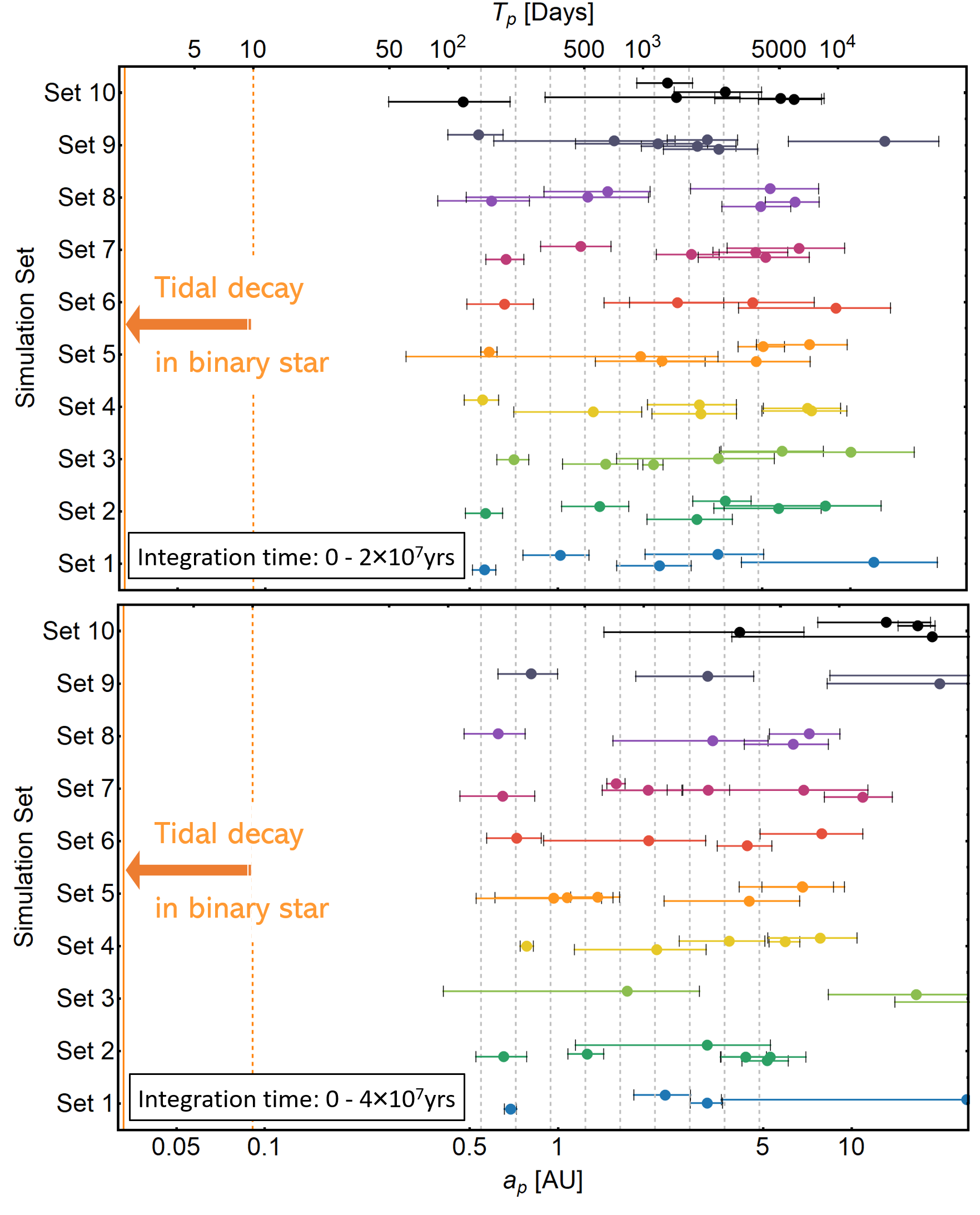}
\end{tabular}
\caption{Architecture of planetary orbits after the tidal evolution of the host binary has ceased.
The results are based on statistical outcomes from 10 runs of the system depicted in Figure \ref{fig:SA nine rings evolution},
each with a random initial value of the longitude of pericenter $\varpi_{p,0}$.
Note that the results in Figure \ref{fig:SA nine rings evolution} correspond to ``Set 6" here.
For each set of numerical integration, the color-coded circles denote the semi-major axes (or orbital periods in days) of the survived planets.
The horizontal line associated with each circle spans the corresponding periastron and apastron distances,
indicating the orbital extent due to nonzero eccentricity.
The gray vertical dashed line marks the initial semi-major axes of the nine planets.
The orange vertical dashed line indicates the initial semi-major axis of the stellar binary $a_{b,0}$,
while the orange vertical solid line shows the binary separation after it has circularized.
The results in the upper and lower panels correspond to different integration time, as labeled.
}
\label{fig:final a and e}
\end{figure}

For the planetary systems of interest here, located within a few AU of the host binary,
the orbital velocity typically exceeds the escape velocity of the planet.
Consequently, $\Theta\lesssim1$ (Equation \ref{eq: Safronov number}),
indicating that a substantial fraction of planetary encounters are expected to result in collisions.
Based on the result in Figure \ref{fig:final a and e},
on average, each simulation produces $\sim3-4$ collision events.
The violent dynamical process significantly reshapes the orbital architecture of the system, manifesting in two features:
\textit{i)} the survival rate of planets in the inner region is extremely low,
with the innermost surviving bodies typically being the products of mergers;
\textit{ii)} planets in the outer region exhibit a higher survival rate,
but their orbits generally migrate outward and retain substantial eccentricities.

It is important to emphasize that even after the stellar binary orbit has circularized,
many planetary orbits retain high eccentricities and continue to cross one another,
as shown in the upper panel of Figure \ref{fig:final a and e}.
This implies that the surviving orbits remain ``unstable",
and their erratic evolution will continue.
Thereby, these planets, which have undergone large-amplitude variation of $e_p$,
are likely to be either ejected from the system or to collide with other CBPs or the central binary stars,
leaving only a very small number of survivors in the inner region
Although the ten realizations are statistically limited,
our simulations already show a clear trend: resonance locking combined with planet–planet scattering efficiently clears planet orbits
\footnote{
To further highlight the role of the stellar binary in triggering the observed instability,
we performed complementary numerical integrations for the same planetary configurations
shown in Figures \ref{fig:SA four rings evolution} and \ref{fig:SA nine rings evolution},
but replaced the binary with a single star of equal total mass and considered only Newtonian gravity.
We find that the four planet system remained stable over $2\times10^7$ yrs,
and the nine planet system showed neither significant eccentricity growth nor system wide disruption.
In contrast, as shown in Figures \ref{fig:SA four rings evolution} and \ref{fig:SA nine rings evolution},
the binary driven secular resonance excites eccentricities rapidly, triggering orbit crossing and scattering on a much shorter timescale. This confirms that binary tidal dissipation and resonance advection are essential for efficient clearing of inner planets.
}.

\section{Conclusion and discussion}
\label{sec 4}

We have studied the evolution and survival of planets around stellar binaries that undergo orbital decay via the tidal dissipation.
An eccentric binary experiencing tidal decay can excite the orbital eccentricities of surrounding planets through apsidal precession resonance.
Once a planet acquires substantial eccentricity, its elongated orbit crosses those of neighboring planets,
triggering planet-planet scatterings.
This process initiates a ``chain reaction", which greatly destabilize the orbits of planets that do not experience eccentricity excitation directly,
leading to planet collision or ejection.
By performing a series of numerical integrations, we show that the mutual gravitational interactions in multi-planet systems
play a major role in amplifying a localized instability due to resonant eccentricity-excitation into a system-wide catastrophe.
Our results suggest that this dynamical process can clear out the inner regions of planetary systems around binaries,
thereby helping to explain the observed ``planetary desert" around compact binaries with orbital period less than $\sim$7 days.
While planet–planet interactions alone can eventually destabilize planetary systems around a single star,
this process typically requires very long timescales \citep[e.g.,][]{Hansen 2020,Wei 2021,Faridani 2022}.
Our study shows that the presence of a stellar binary dramatically accelerates this instability in multi-planet systems.

Throughout this \textit{Letter}, we have adopted a representative set of parameters to illustrate the dynamical process.
While our numerical examples assume a fiducial binary initial eccentricity of $e_{b,0}=0.8$,
the resonance-induced instability can occur for systems with $e_{b,0}\gtrsim0.5$
(a threshold that remains robust for different tidal timescales; see Appendix \ref{Appendix A}).
A lower initial binary eccentricity would restrict the range of planetary orbits susceptible to eccentricity excitation
and prolong the timescale of stellar tidal decay.
In general, once a planet acquires significant eccentricity excitation,
planet-planet interactions would spread the instability to a wide region of the system.
In our study, we have considered a fiducial binary configuration and assumed uniform planetary masses.
These choices set the characteristic timescale for eccentricity excitation.
Varying the planet mass distribution would modify the Hill radii and scattering cross sections,
while a larger radius would increase collision probabilities.
These changes would modify the efficiency of dynamical clearing.
A systematic exploration across a broader parameter space lies beyond the scope of the present study and is reserved for future work.

To fully explain the observational statistics of circumbinary planets,
future population synthesis simulations that couple binary tidal evolution and the dynamical processes identified here is essential.
Such simulations will enable quantitative predictions of planet occurrence rates around binaries of different periods.

\section{Acknowledgments}

B. L. thanks Yanqin Wu, Douglas N.C.Lin and Yihan Wang for useful discussions.
B. L. acknowledges support from the National Natural Science Foundation of China (Grant No. 12433008)
and National Key Research and Development Program of China (No. 2023YFB3002502).

\appendix
\section{A: Dependence on the Tidal Dissipation Timescale}
\label{Appendix A}

In the main text, we adopt the equilibrium tidal model with a constant lag time,
a widely used prescription in the literature \citep[e.g.,][]{Hut1981,Eggleton2001,Fabrycky2007}.
This parameterized model captures the essential effect of binary orbital decay.
In our scenario, the key condition for successful resonance capture and advection is that the binary decay timescale be sufficiently long
compared to the apsidal precession timescale of the outer orbit;
that is, the evolution must be adiabatic.
This condition does not depend on the specific tidal model.
Any mechanism that drives slow orbital decay, such as gravitational wave emission, magnetic braking, or tidal dissipation,
can potentially trigger the resonance.
To test the sensitivity of our results to the assumed decay rate,
we present in this Appendix a comparison with simulations using a longer tidal timescale $t_V =50$ yr.

\begin{figure}
\centering
\begin{tabular}{cccc}
\includegraphics[width=8.5cm]{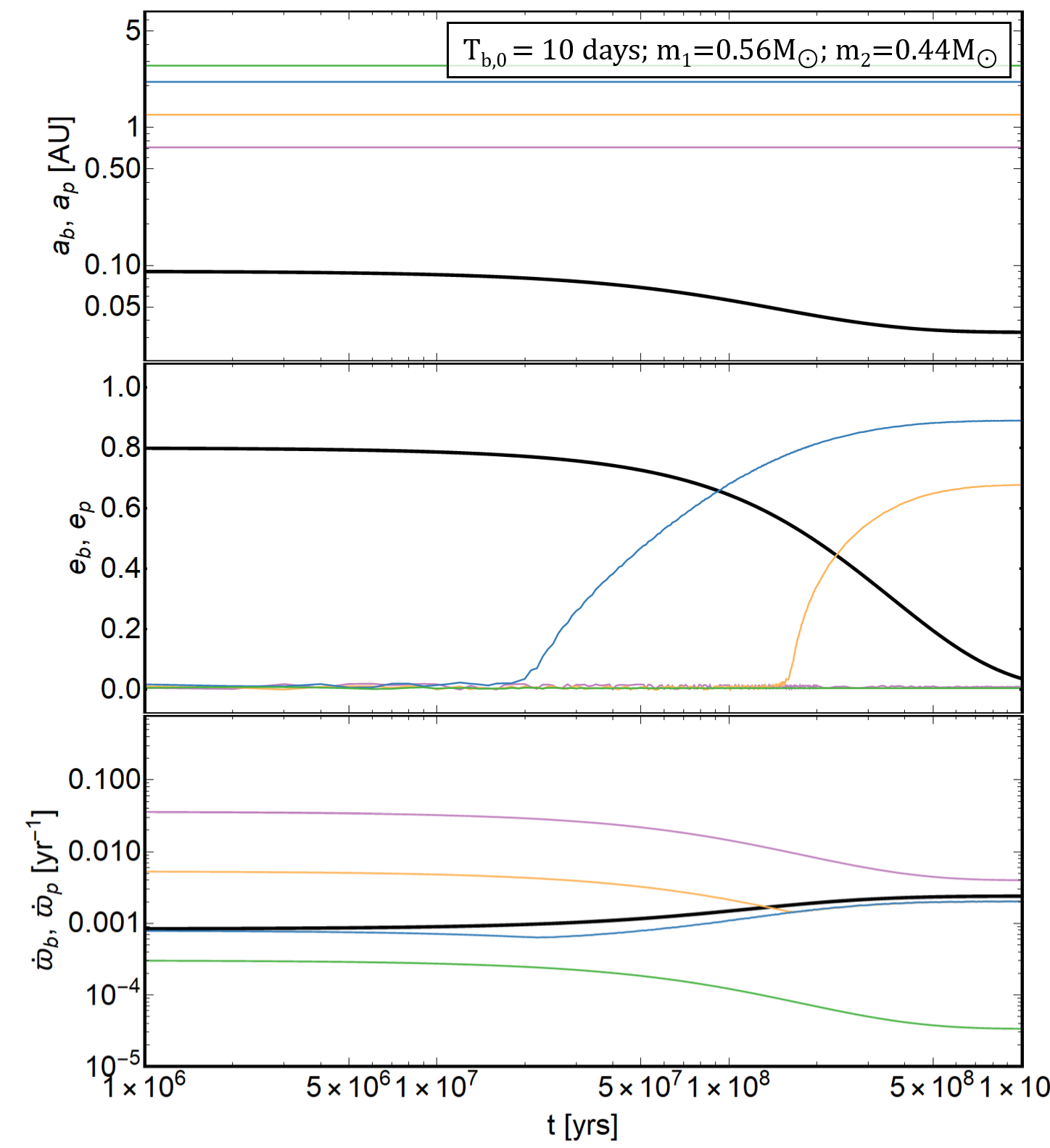}
\end{tabular}
\caption{Similar to Fig. \ref{fig:SA evolution} in the main text, but with $t_V =50$ yr.
}
\label{fig:app_tV_comparison}
\end{figure}

Figure \ref{fig:app_tV_comparison} shows the evolution of a stellar binary (black)
with four non interacting circumbinary planets in a coplanar configuration.
Compared to the case in Figure \ref{fig:SA evolution},
the planet can be captured into resonance advection and experiences significant eccentricity growth in both cases.
The qualitative behavior remains unchanged, where the apsidal precession rates lock, leading to extreme eccentricity excitation.
The main difference lies in the timescale.
For $t_V =50$, the binary decays more slowly, and resonance capture occurs later.
Nevertheless, the final eccentricity reaches a comparable level.

\begin{figure}
\centering
\begin{tabular}{cccc}
\includegraphics[width=8.5cm]{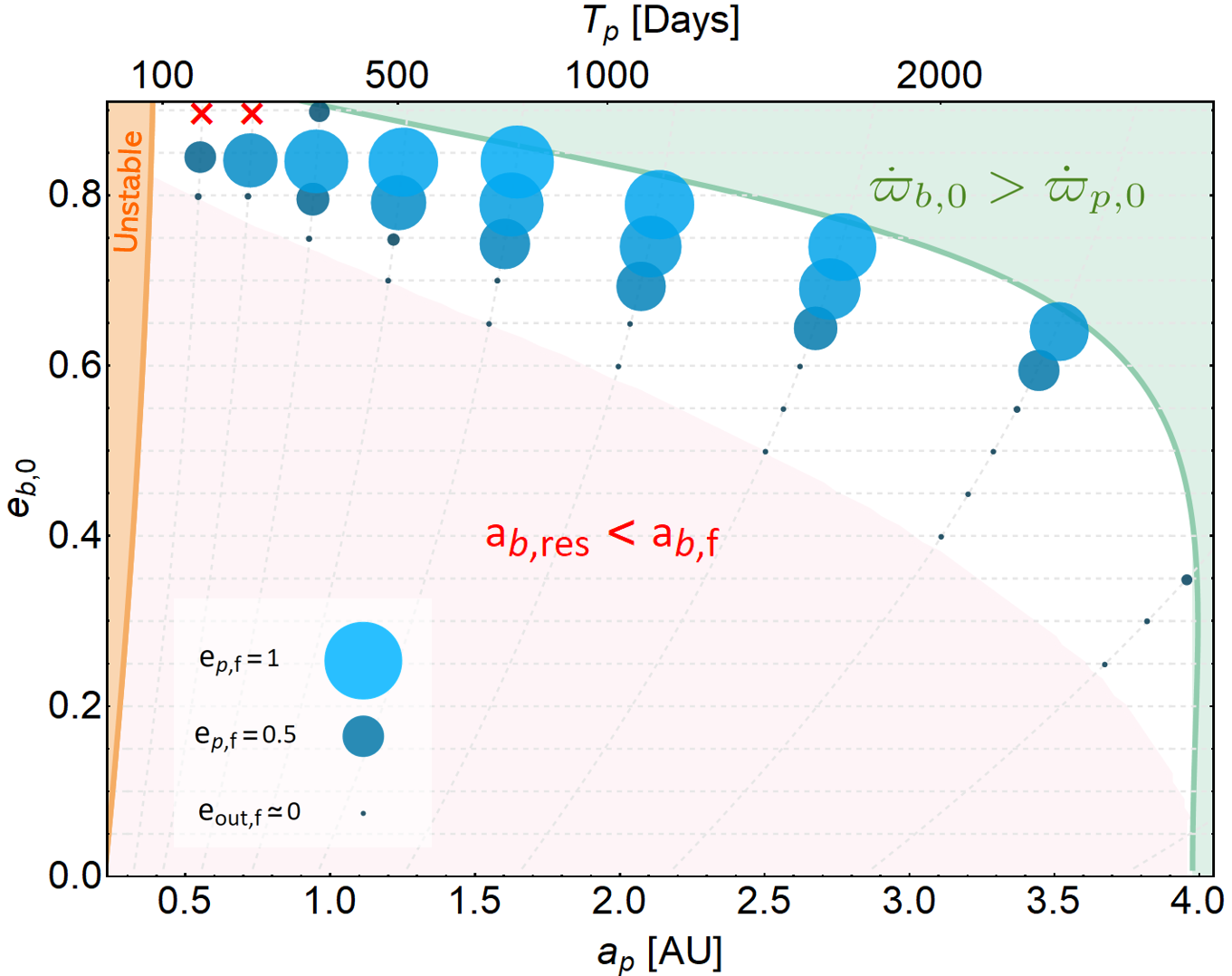}
\end{tabular}
\caption{Similar to Fig. \ref{fig:parameter space} in the main text, but with $t_V =50$ yr.
}
\label{fig:app_tV_comparison P}
\end{figure}

Figure \ref{fig:app_tV_comparison P} presents the same parameter space as Figure \ref{fig:parameter space}
but for a longer tidal dissipation timescale $t_V =50$ yr.
The overall structure remains similar.
However, because the binary decays more slowly,
the resonance capture condition becomes more stringent.
For binaries with initial eccentricity $e_{b,0}\lesssim 0.7$, the tidal decay does not complete within $10^{10}$ yr.
Consequently, in systems where $\dot\varpi_p$ would have overtaken $\dot\varpi_b$ in the fiducial case ($t_V =1$ yr),
the overtaking fails to occur before the integration stops ($t_V =50$ yr).
As a result, for binaries with $e_{b,0}\lesssim 0.6$, the planetary eccentricity is not excited.

Based on the results, we conclude that our results are robust against reasonable variations in the tidal dissipation timescale.
The core mechanism of resonance driven eccentricity excitation
followed by planet–planet clearing does not depend sensitively on the precise value of $t_V $, provided the binary decay remains adiabatic.

Note that the systems considered in the main text, with initial eccentricity $e_{b,0}=0.8$, are highly eccentric throughout much of the evolution.
The dynamical tides may operate in such systems \citep[e.g.,][]{Fuller 2012,Vick 2019},
but we do not expect it to qualitatively affect our conclusions,
as the resonance condition relies primarily on adiabatic decay rather than the specific tidal model.

%%%%%%%%%%%%%%%%%%%%%%%%%%%%%%%%%%%%%%%%%%%%%%%%%%%%%%%%%%%%%%%%%%%%%%

\end{document}